\documentclass[aps,prb,twocolumn,superscriptaddress,showpacs]{revtex4}
\usepackage{graphicx}

\begin{document}

\title{Longitudinal versus transverse spheroidal vibrational
modes of an elastic sphere}

\author{Lucien Saviot}
\affiliation{Laboratoire de Recherche sur la R\'eactivit\'e des Solides,
UMR 5613 CNRS - Universit\'e de Bourgogne\\
9 avenue A. Savary, BP 47870 - 21078 Dijon - France}
\email{lucien.saviot@u-bourgogne.fr}

\author{Daniel B. Murray}
\affiliation{Department of Physics and Astronomy,
The University of British Columbia Okanagan, 3333 University Way,
Kelowna, British Columbia, Canada V1V 1V7}
\email{daniel.murray@ubc.ca}

\date{\today}

\begin{abstract}
Analysis of the spheroidal modes of vibration of a free
continuum elastic sphere show that they can be qualitatively grouped
into two categories: primarily longitudinal and primarily
transverse. This is not a sharp distinction.  However,
there is a relatively stark contrast between the two kinds
of modes.  Primarily transverse modes have a small divergence
and have frequencies that are almost functionally independent
of the longitudinal speed of sound.  Analysis of inelastic
light scattering intensity from confined acoustic phonons
in nanoparticles requires an understanding of this qualitative
distinction between different spheroidal modes. In
addition, some common misconceptions about spheroidal modes
are corrected.
\end{abstract}

\maketitle

\section{Introduction}
\label{Introduction}

With the explosion of interest in the optical properties of
nanoparticles, the classic elastic mechanical problem of
the vibrational modes of a free continuum sphere has found
a new context for application.  The problem was formally
and numerically solved back in 1882.\cite{lamb1882}
Nanoparticles, \textit{i.e.} spherical clusters of atoms ranging
in diameter from 1~nm to 100~nm, have sufficiently few atoms
that the continuum approximation can be
questioned.\cite{ChengPRB05,ErratumChengPRB05}
Even so, it is acceptable to ignore the effects of the discrete
crystal lattice for the few vibrational modes with lowest
frequency as long as the nanoparticle diameter exceeds
several nanometers.

Inelastic light scattering of a continuous laser beam shining on
the nanoparticle permits detection of the mechanical vibrations
since the changing size and shape of the nanoparticle modulates
the polarizability of the nanoparticle, so that the monochromatic
incident light turns into scattered light with sidebands shifted
up and down by the frequency of the vibrations.  This can be
seen using experimental setups of the Raman and Brillouin type.

For theoretical convenience the material is assumed to be
homogeneous, isotropic and linear.  The outer surface of the
sphere is free from externally imposed stresses and this
situation will be referred to as the ``free sphere model'' (FSM).
The original paper by Lamb\cite{lamb1882} classified the FSM
modes of vibration into two classes, now called ``torsional''
(TOR) and ``spheroidal'' (SPH).  The distinctive feature of
torsional modes is that the material density does not vary.  In
other words, the divergence of the displacement field is zero.
Furthermore, the spherical symmetry permits classification of
the modes by angular momentum number $\ell \geq 0$.
(However, later on we will show that there is value
in considering $\ell$ to be a continuous variable.)  There is no
dependence of the frequency on the $z$ angular momentum $m$.
Beyond this, the modes are indexed by $n \geq 0$.
It is convenient to let $p$ denote either SPH or TOR,
to indicate individual modes by $(p,\ell,n)$ and their
frequencies by $\omega_{p \ell n}$.

What we explore in this paper is an additional classification of
the SPH modes beyond that employed since Lamb.  In particular,
SPH modes can be classified (albeit approximately and
subjectively) as either being primarily longitudinal (SPH$_L$)
or primarily transverse (SPH$_T$) in nature. The specific meaning
of this will be explained further on.  This is not a sharp
division, and actual modes fall somewhere in between the two
ideals.  However, the contrast is sufficiently sharp that this
new distinction among SPH FSM modes as SPH$_L$ or SPH$_T$ is a
very important tool.
                                                  
In a recent theoretical paper, G.~Bachelier and
A.~Mlayah\cite{bachelierPRB04} predicted that
(SPH,$\ell=2$,$n$) modes with differing values of $n$ 
contribute to the Raman spectrum in a highly non-uniform way.
In this paper we will show that this can be explained using
the previously mentionned distinction between SPH modes.
They pointed out that there are two separate mechanisms that
couple (SPH,$\ell=2$) acoustic vibrations to the surface plasmon
resonance and in turn lead to Raman scattering.  First, change
of the particle shape and second, modulation of the density
leading to change of optical properties through the deformation
potential.

Section \ref{Freesphere} reprises the formalism necessary for the       
FSM solution. In Section \ref{Longitudinality}, we show explicitly      
what we mean by SPH$_L$ and SPH$_T$. In Section \ref{Highfreq}, we      
illustrate the natural appearance of SPH$_L$ and SPH$_T$ modes in the   
high frequency limit. Section \ref{Discussion} discusses these results  
and their connection with inelastic light scattering experiments.       

\section{The Free Sphere Model}
\label{Freesphere}

Vibrational modes of a free linear elastic continuum homogeneous
isotropic sphere were found by Lamb in 1882.\cite{lamb1882}

For a mode with angular frequency $\omega$, the displacement of
material point $\vec{r}$ from its equilibrium position is
$\vec{u}(\vec{r}) cos(\omega t)$.  For a $m=0$ TOR mode, 
$\vec{u}$ =
A $\nabla \times ( \vec{r} j_{\ell}(k_T r) P_{\ell}(cos \theta) )$,
where $j_{\ell}$ are spherical Bessel functions of the first kind
and $P_{\ell}$ are Legendre polynomials.  For a $m=0$ SPH mode,
$\vec{u}$ = $\vec{u}_L$ + $\vec{u}_T$ where
\begin{equation}
\vec{u}_L(r,\theta) = B \nabla j_{\ell}(k_L r) P_{\ell}(cos \theta) 
\label{Bterm}
\end{equation}
and
\begin{equation}
\vec{u}_T(r,\theta) = C \nabla \times \nabla \times ( \vec{r} j_{\ell}(k_T r) P_{\ell}(cos \theta) )
\label{Cterm}
\end{equation}
where $A$, $B$ and $C$ are real coefficients,
$v_L k_L$ = $v_T k_T$ = $\omega$,
and $v_T$ and $v_L$ are the transverse and longitudinal speeds
of sound.

Modes with $z$ angular momentum $m \ne 0$ have a different
functional form. 

$R$ is the nanoparticle radius.  If $\sigma_{ij}$ is the stress
tensor, the boundary conditions at $r$ = $R$ are $\sigma_{rr} $
= $\sigma_{r \theta}$ = 0.  It is convenient to introduce
dimensionless frequencies $\eta$ = $k_T R$ and $\xi$ = $k_L R$.
Following Eringen,\cite{eringen} application of these boundary
conditions determines the allowed SPH vibrational frequencies as
zeroes of a 2 $\times$ 2 determinant for $\ell>0$.
\begin{equation}
  \Delta_{\ell} =
	\begin{array}{|cc|}
	  T_{11} & T_{13}\\
		T_{41} & T_{43}
	\end{array}
\label{twobytwo}
\end{equation}
where
\begin{eqnarray}
\nonumber       T_{11} & = &
        \left(\ell^2 - \ell - \frac{\eta^2}2\right) j_{\ell}(\xi) + 2 \xi j_{\ell+1}(\xi)\\
\nonumber	T_{13} & = &
        \ell (\ell + 1)\left\{ (\ell - 1) j_{\ell}(\eta) - \eta
        j_{\ell+1}(\eta)\right\}\\
\nonumber	T_{41}& = &
        (\ell - 1) j_{\ell}(\xi) - \xi j_{\ell+1}(\xi)\\
\nonumber	T_{43} & = &
        \left(\ell^2 - 1 - \frac{\eta^2}2 \right) j_{\ell}(\eta) + \eta j_{\ell+1}(\eta)
\end{eqnarray}

For $\ell=0$, the allowed vibrational frequencies are the zeroes
of $T_{11}$.

Noting that the displacement fields are real-valued, it is
appropriate to use the following inner product between two
displacement fields $u_A$ and $u_B$:\cite{murrayPRB04}

\begin{equation}
(u_A | u_B) =
\frac{\int_{r<R} \vec{u_A}(\vec r) \cdot \vec u_B(\vec r) \rho d^3\vec{r}}
     {\int_{r<R} \rho d^3\vec{r}}
  \label{norm}
\end{equation}

\begin{figure}[!ht]
  \includegraphics[width=\columnwidth]{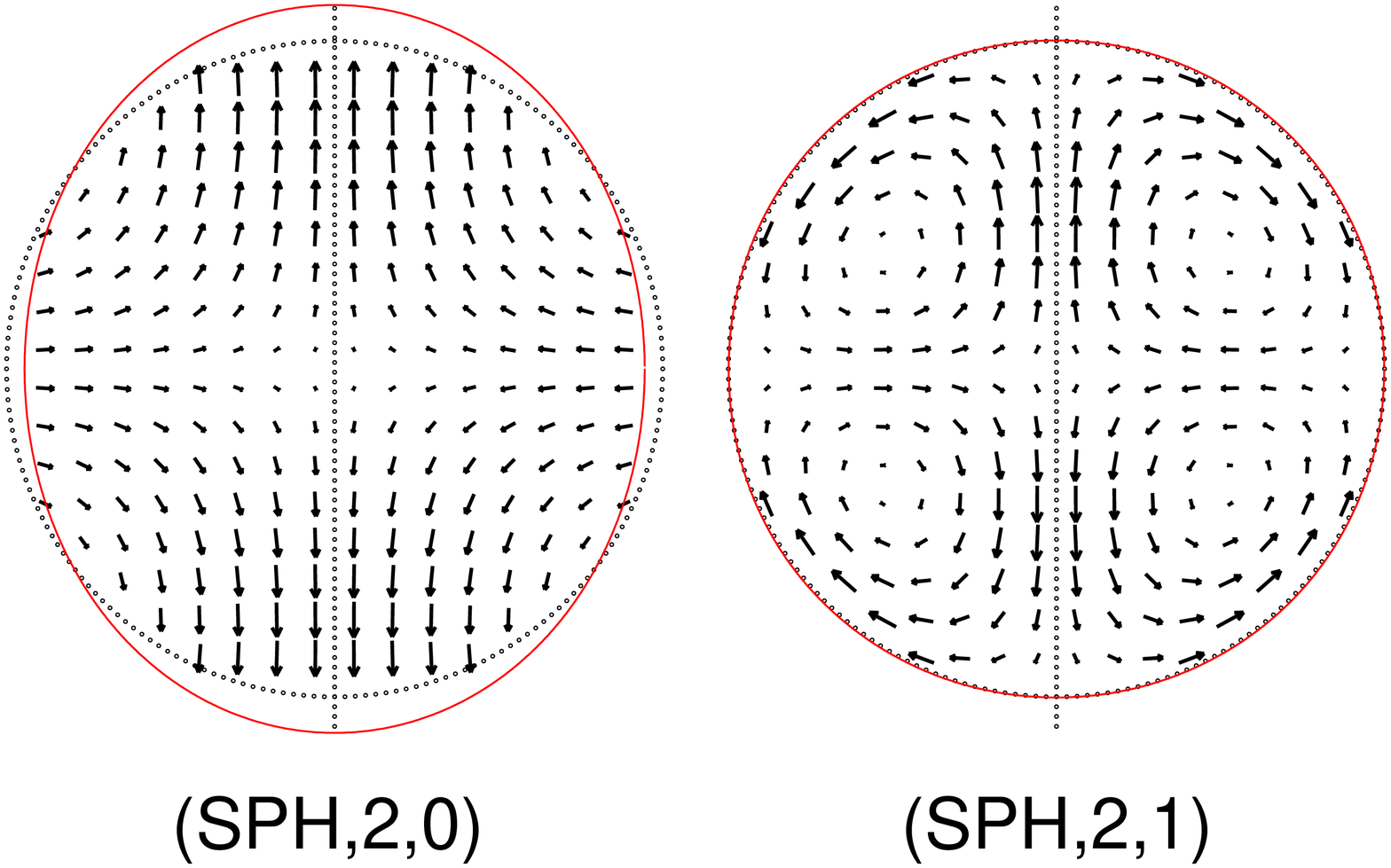}
  \includegraphics[width=\columnwidth]{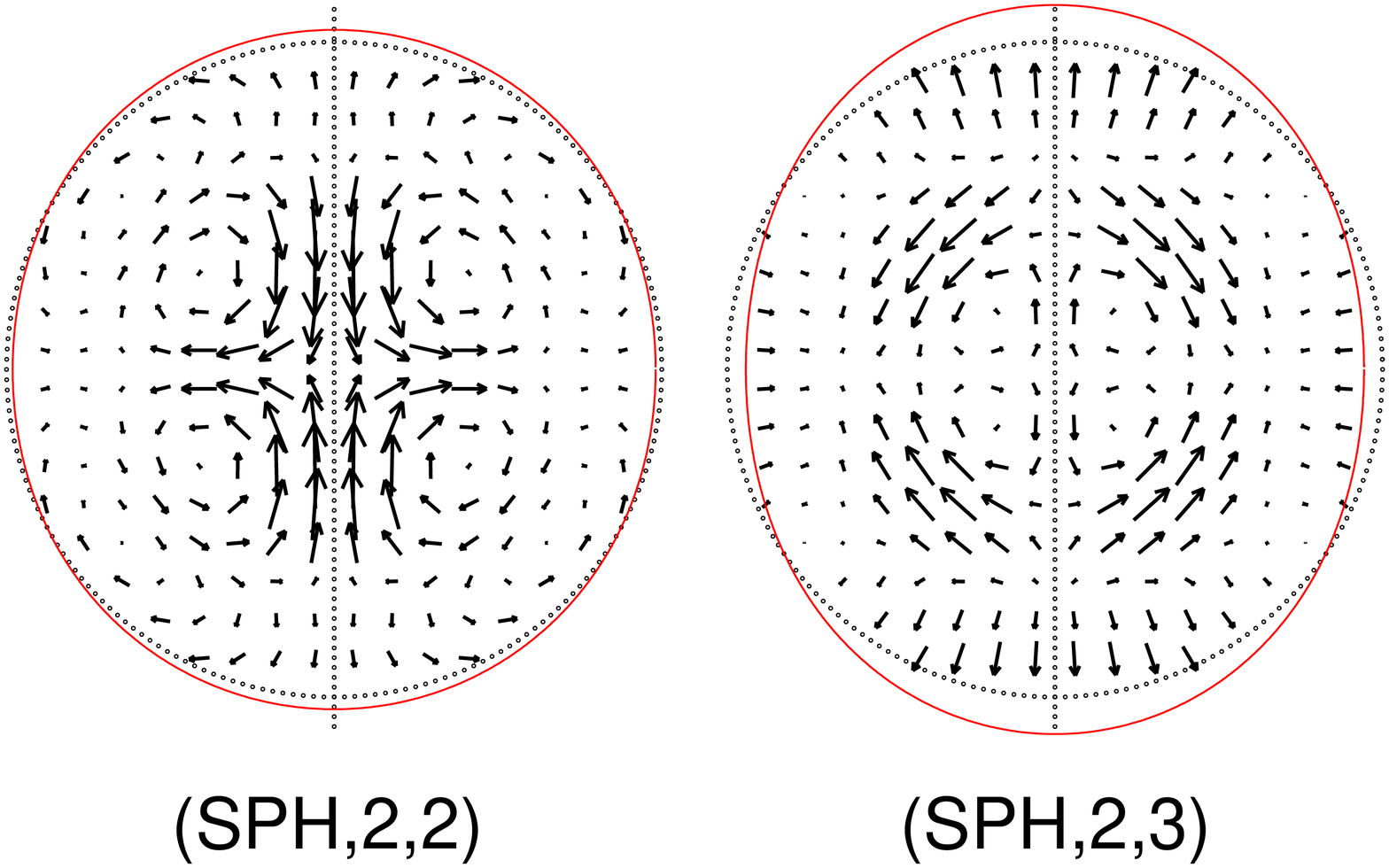}
  \caption{\label{prf12}(Color online) Displacement fields
  $\vec{u}(\vec{r})$ of selected SPH $\ell$=2 modes. As explained in the
  text, the first three are primarly transverse (\textit{i.e.} SPH$_T$).
  (SPH,2,3) is primarly longitudinal (\textit{i.e.} SPH$_L$). The
  equilibrium surface of the nanoparticle and the $z$-axis are shown as
  dotted lines. The solid (red online) line shows the distorted surface.
  Note that the (SPH,2,1) mode does not change the nanoparticle shape.}
\end{figure}

A normalization condition (such as ($u$$|$$u$) = 1) would
typically determine the final values of $B$ and $C$.  But the
details of the condition do not affect the results reported
here.  The displacement field $\vec{u}(\vec{r})$ for some
selected modes are depicted in Fig.~\ref{prf12}.

\section{Spheroidal Mode Longitudinality}
\label{Longitudinality}

\begin{figure}[!ht]
 \includegraphics[width=\columnwidth]{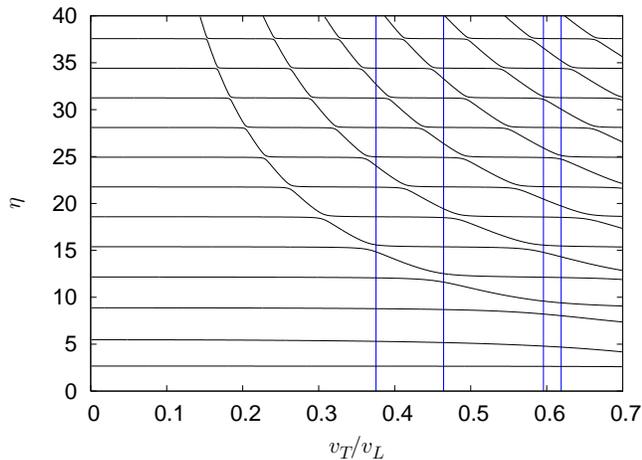}
 \caption{\label{SPH2}(Color online) Dimensionless mode frequency $\eta$
 as a function of $v_T/v_L$ for (SPH,$\ell=2$) modes. Vertical lines
 (blue online) mark $v_T/v_L$ for Au, Ag, Si and Ge from left to right.}
\end{figure}

Isotropic elastic materials differ in their Poisson ratio,
$\nu$, which is related to $x = v_T/v_L$ through
$x = \sqrt{(1 - 2 \nu)/(2 - 2 \nu)}$.  Figure~\ref{SPH2} shows how
the dimensionless frequency, $\eta$, of the SPH $\ell = 2$ FSM
modes varies with $v_T/v_L$.  It is quite apparent that some
modes keep the same $\eta$ as $v_T/v_L$ is varied.  However,
other modes change frequency as $v_T/v_L$ changes.  There
are transition points where a given mode changes from
being constant to varying with $v_T/v_L$.

This pattern visible in Fig.~\ref{SPH2} motivates the search
for a numerical criterion to permit this contrast among modes to
be quantified. We adopt the starting point that in some
sense some modes are more transverse in nature (SPH$_T$) while others
are more longitudinal (SPH$_L$).  We then coin the term ``longitudinality'',
denoted by $L$, for a quantity that varies on a scale from 0 to
1 with 0 being purely transverse and 1 being purely longitudinal.
There is no single obvious way of doing this.  Rather, we have
evaluated a number of quantities as candidates for the best
measure of longitudinality, of which we present four which
work well.  These will be denoted $L1$, $L2$, $L3$, and $L4$.

Consider a particular SPH mode with indices $\ell$ and $n$. Its         
frequency is $\omega(v_L,v_T)$. Define $L1$ by                          
\begin{equation}
L1 = \frac{v_L}{\omega}  \frac{\partial \omega}{\partial v_L}
   = - \frac{x}{\eta}  \frac{d \eta}{d x}
   = 1 - \frac{v_T}{\omega}  \frac{\partial \omega}{\partial v_T}
\label{eqL}
\end{equation}

Noting that
$\vec{u}(\vec{r}) = \vec{u}_T(\vec{r}) + \vec{u}_L(\vec{r})$,
we define
$L2$ as $(u_L|u_L) / (u|u)$,
and $L3$ as $1 - \left((u_T|u_T) / (u|u)\right)$.
But note also that
$(u_L|u_L) + (u_T|u_T) \ne (u|u)$
since $(u_L|u_T) \ne 0$.

\begin{figure}[!ht]
  \includegraphics[width=\columnwidth]{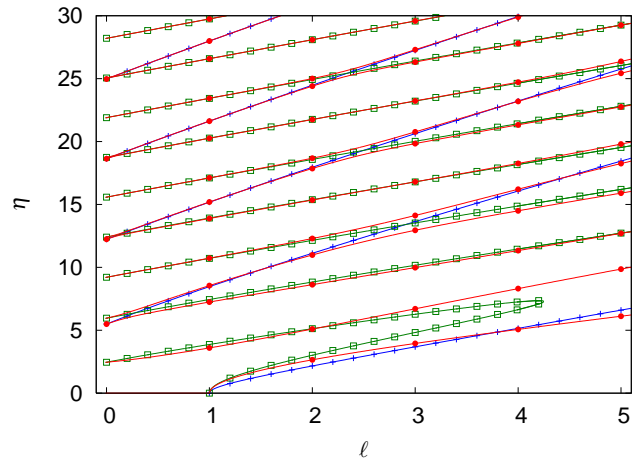}
  \caption{\label{etavsl}(Color online) Variation of the SPH mode
  frequency with $\ell$ for a material with $v_T/v_L$ = 0.5. Full
  circles are exact FSM frequencies and are connected with curves
  calculated for non-integer $\ell$ (red online). Lines with crosses
  are roots of $T_{11}$ (blue online) which approximate SPH$_L$ modes.
  Lines with empty squares are roots of $T_{43}$ (green online) which
  approximate SPH$_T$ modes.}
\end{figure}

Given a fixed value of $v_T/v_L$ and $n$, $\eta$ may be considered
to be a continuous function of $\ell$, as in Fig.~\ref{etavsl}.
In terms of this $\eta(\ell)$, define
\begin{equation}
L4 = \frac{v_T}{v_L-v_T}
       \left( \frac2\pi \frac{d\eta}{d\ell} - 1 \right)
\end{equation}

Let $<..>_V$ and $<..>_S$ denote averages over the nanoparticle
volume and surface, respectively. In particular,
$(u|u) = < u_r^2 + u_\theta^2 + u_\phi^2 >_V$.
Some other measures of interest are as follows:
URV = $< u_r^2 >_V/(u|u)$,
URS = $< u_r^2 >_S/(u|u)$,
UTS = $< u_\theta^2 + u_\phi^2 >_S/(u|u)$,
and U2S = URS + UTS.

Note that all of these quantities are defined in such a
way as to be independent of $m$.

\begin{figure}[!ht]
  \includegraphics[width=\columnwidth]{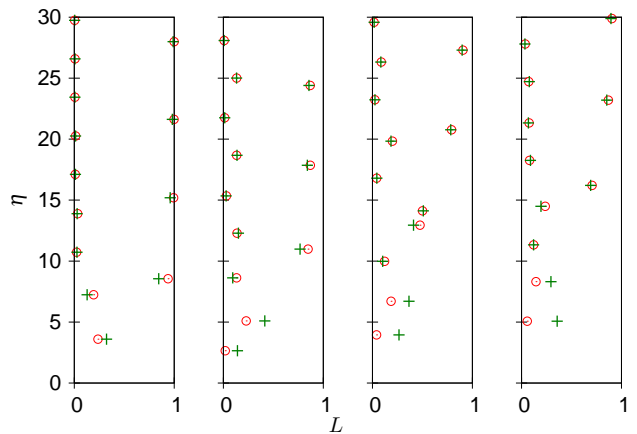}
  \caption{\label{Poisson2}(Color online) Dimensionless mode frequency
  $\eta$ as a function of longitudinality measures $L1$ (circles (red
  online)) and $L2$ (crosses (green online)) for SPH modes of a material
  with $v_T/v_L=0.5$. $\ell = 1 \ldots 4$ from left to right. SPH$_T$
  modes have L1 and L2 near zero, while for SPH$_L$ modes they are close
  to 1.}
\end{figure}

\begin{table}
\caption{\label{AllL}Longitudinality measures for a material with
$v_T$/$v_L$ = 0.5.}
\begin{tabular}{|c|c|c|c|c|c|c|c|c|c|c|}
\toprule
$\ell$&$n$&$\eta$ & L1 & L2 & L3 & L4\\
\colrule
0 & 0 &  5.49 & 1.36 & 1.00 & 1.00 & 0.86\\
0 & 1 & 12.23 & 1.06 & 1.00 & 1.00 & 0.84\\
0 & 2 & 18.63 & 1.02 & 1.00 & 1.00 & 0.89\\
\colrule
1 & 0 &  3.60 & 0.24 & 0.32 &-0.27 &-0.12\\
1 & 1 &  7.24 & 0.19 & 0.13 & 0.10 &-0.09\\
1 & 2 &  8.55 & 0.94 & 0.84 & 0.84 & 0.66\\
1 & 3 & 10.73 & 0.02 & 0.03 & 0.02 &-0.03\\
1 & 4 & 13.89 & 0.03 & 0.03 & 0.03 &-0.06\\
1 & 5 & 15.19 & 0.99 & 0.96 & 0.96 & 0.79\\
1 & 6 & 17.11 & 0.01 & 0.01 & 0.01 &-0.03\\
\colrule
2 & 0 &  2.65 & 0.02 & 0.14 &-0.85 &-0.06\\
2 & 1 &  5.10 & 0.23 & 0.41 & 0.02 & 0.01\\
2 & 2 &  8.63 & 0.13 & 0.09 & 0.04 &-0.13\\
2 & 3 & 10.99 & 0.85 & 0.77 & 0.75 & 0.42\\
2 & 4 & 12.29 & 0.14 & 0.15 & 0.15 & 0.05\\
2 & 5 & 15.35 & 0.03 & 0.03 & 0.02 &-0.07\\
2 & 6 & 17.85 & 0.87 & 0.84 & 0.84 & 0.57\\
2 & 7 & 18.68 & 0.13 & 0.14 & 0.14 & 0.07\\
\colrule
3 & 0 &  3.95 & 0.04 & 0.27 &-1.15 &-0.25\\
3 & 1 &  6.71 & 0.19 & 0.37 & 0.24 & 0.03\\
3 & 2 &  9.98 & 0.12 & 0.10 & 0.04 &-0.14\\
3 & 3 & 12.95 & 0.48 & 0.41 & 0.38 & 0.08\\
3 & 4 & 14.12 & 0.50 & 0.51 & 0.51 & 0.28\\
3 & 5 & 16.80 & 0.04 & 0.04 & 0.04 &-0.07\\
3 & 6 & 19.84 & 0.20 & 0.19 & 0.18 & 0.01\\
\colrule
4 & 0 &  5.07 & 0.06 & 0.36 &-1.30 &-0.31\\
4 & 1 &  8.31 & 0.14 & 0.29 & 0.33 & 0.01\\
4 & 2 & 11.33 & 0.12 & 0.12 & 0.06 &-0.14\\
4 & 3 & 14.49 & 0.24 & 0.19 & 0.16 &-0.07\\
4 & 4 & 16.21 & 0.71 & 0.69 & 0.69 & 0.33\\
4 & 5 & 18.26 & 0.08 & 0.09 & 0.09 &-0.05\\
\botrule
\end{tabular}
\end{table}

   Except at low $\eta$, Fig.~\ref{Poisson2} shows that L1
and L2 are in close agreement. L3 and L4 are not plotted, but
also agree closely except at low $\eta$ (see Table~\ref{AllL}).
Generally, a given mode either
has all of L1, L2, L3, and L4 low, or else all high.
It is thus possible
to classify modes as SPH$_L$ or SPH$_T$.  Rarely, there are
cases where the values of L1, L2, L3, and L4 are in the intermediate
range, such as in Fig.~\ref{Poisson2} for $\ell = 3$ for
the two modes near $\eta = 13$.  Such modes which are neither
clearly SPH$_L$ nor SPH$_T$ always occur in pairs.
The reason for this is explained in Section~\ref{Highfreq}.

Table~\ref{def} provides additional information about the modes.
The dimensionless frequency $\eta$ is provided for convenience,
as is the ratio of coefficients $B$ and $C$.

In principle, $C/B$ could be expected to provide useful
information about whether a mode is SPH$_L$ or SPH$_T$.  In the
extreme case that $C = 0$, the mode is evidently SPH$_L$,
and likewise when $B = 0$ the mode is SPH$_T$.  But the
$C/B$ values do not exhibit an informative pattern.

There is a strong contrast in the values of URS for
different modes.  However, it does not correlate to whether the
mode is SPH$_L$ or SPH$_T$ except at high enough $\eta$.
URS is an interesting quantity
because it is the one we have to monitor for the surface
deformation mechanism except for $\ell=0$ modes.

\begin{table}
\caption{\label{def}Other features of SPH modes for a material with
$v_T$/$v_L$ = 0.5.}
\begin{tabular}{|c|c|c|c|c|c|c|c|c|c|c|}
\toprule
$\ell$&$n$&$\eta$ &$C/B$&URV &  URS &  UTS &  U2S & Class  &$n_L$&$n_T$\\
\colrule
0 & 0 &  5.49 &  0.00 & 1.00 & 0.91 & 0.00 & 0.91 & SPH$_L$ & 0 &  \\
0 & 1 & 12.23 &  0.00 & 1.00 & 0.71 & 0.00 & 0.71 & SPH$_L$ & 1 &  \\
0 & 2 & 18.63 &  0.00 & 1.00 & 0.68 & 0.00 & 0.68 & SPH$_L$ & 2 &  \\
\colrule
1 & 0 &  3.60 & -0.99 & 0.37 & 0.05 & 1.26 & 1.31 & SPH$_T$ &   & 0 \\
1 & 1 &  7.24 &  1.65 & 0.43 & 0.14 & 0.66 & 0.81 & SPH$_T$ &   & 1 \\
1 & 2 &  8.55 & -0.29 & 0.43 & 0.70 & 0.03 & 0.72 & SPH$_L$ & 0 &   \\
1 & 3 & 10.73 &  4.37 & 0.21 & 0.00 & 0.69 & 0.69 & SPH$_T$ &   & 2 \\
1 & 4 & 13.89 & -3.93 & 0.14 & 0.02 & 0.68 & 0.69 & SPH$_T$ &   & 3 \\
\colrule                                                   
2 & 0 &  2.65 & -0.44 & 0.59 & 0.66 & 0.20 & 0.86 & SPH$_T$ &   & 0 \\
2 & 1 &  5.10 & -0.38 & 0.26 & 0.00 & 1.78 & 1.78 & SPH$_T$ &   & 1 \\
2 & 2 &  8.63 &  1.09 & 0.50 & 0.08 & 0.85 & 0.92 & SPH$_T$ &   & 2 \\
2 & 3 & 10.99 & -0.22 & 0.35 & 0.71 & 0.04 & 0.75 & SPH$_L$ & 0 &  \\
2 & 4 & 12.29 &  0.96 & 0.34 & 0.07 & 0.64 & 0.71 & SPH$_T$ &   & 3\\
\colrule
3 & 0 &  3.95 & -0.17 & 0.74 & 0.89 & 0.03 & 0.91 & SPH$_T$ &   & 0 \\
3 & 1 &  6.71 & -0.21 & 0.22 & 0.04 & 1.91 & 1.95 & SPH$_T$ &   & 1 \\
3 & 2 &  9.98 &  0.70 & 0.53 & 0.04 & 1.05 & 1.09 & SPH$_T$ &   & 2 \\
3 & 3 & 12.95 & -0.33 & 0.35 & 0.43 & 0.41 & 0.84 & mix     & 0 & 3 \\
3 & 4 & 14.12 &  0.27 & 0.37 & 0.39 & 0.29 & 0.68 & mix     & 0 & 3 \\
\colrule                                          
4 & 0 &  5.07 & -0.08 & 0.81 & 1.05 & 0.00 & 1.05 & SPH$_T$ &   & 0 \\
4 & 1 &  8.31 & -0.15 & 0.23 & 0.14 & 1.78 & 1.92 & SPH$_T$ &   & 1 \\
4 & 2 & 11.33 &  0.46 & 0.54 & 0.01 & 1.25 & 1.27 & SPH$_T$ &   & 2 \\
4 & 3 & 14.49 & -0.41 & 0.41 & 0.21 & 0.75 & 0.95 & SPH$_T$ &   & 3 \\
4 & 4 & 16.21 &  0.14 & 0.35 & 0.65 & 0.03 & 0.68 & SPH$_L$ & 0 &   \\
\botrule
\end{tabular}
\end{table}

Group theoretical arguments\cite{duval92} show that only
SPH modes with $\ell$=0 and $\ell$=2 are Raman active.
This assumes that the nanoparticle is perfectly spherical
in shape and spherically symmetric in all of its properties.
The basic nature of the $\ell=0$ modes is much more clear
because of their simplicity and symmetry.  Consequently,
the modes (SPH,$\ell=2$,$n$) are of primary interest when
trying to understand Raman intensities.

From the value of L2 $\simeq$ 0.14 in Tab.~\ref{def}, the displacement  
of (SPH,2,0) is mostly due to the $u_T$ term and not the $u_L$ term.    
Its squared displacement due to the $u_L$ term alone is  
just 14\% of the total. The $u_T$ term has zero divergence. Therefore,  
(SPH,2,0) doesn't have much divergence. So the effect of changing       
density on the dielectric constant through the deformation potential    
may not be significant to the overall Raman intensity.                  

On the other hand, based on its URS of $\simeq$ 0.66 and UTS
of $\simeq$ 0.20, the surface displacement of (SPH,2,0)
is strongly along $r$ and only weakly along $\theta$ as
Fig.~\ref{prf12} illustrates.  Note that, $r$ surface
displacement changes the nanoparticle shape, while $\theta$
displacement does not.

The (SPH,2,1) mode differs from (SPH,2,0) in several ways. From the     
$L1$ value of 0.2281 in Fig.~\ref{Poisson2}, we can see that the        
frequency of (SPH,2,1) depends more on $v_L$. Also, $L2 \simeq 0.416$   
in Fig.~\ref{Poisson2} shows that (SPH,2,1) has more of a $u_L$         
component, even if it is still weaker than the $u_T$ part. But this     
means that (SPH,2,1) can have much more divergence than (SPH,2,0).      
So the deformation potential mechanism can modulate the dielectric      
constant. But it is very interesting to notice from the URS value of    
$\simeq$ 0.00 in Tab.~\ref{def} (more precisely, 0.0003) that (SPH,2,1) 
causes negligible radial movement of the surface. So (SPH,2,1) barely   
changes the shape of the nanoparticle, as Fig.~\ref{prf12} shows.       

(SPH,2,3) has strong $v_L$
dependence ($L1 \simeq 0.8475$) in Fig.~\ref{Poisson2} as well
as a strong $u_L$ component ($L2 \simeq 0.766$).  So it is clear
that it is SPH$_L$.   It's surface displacement is mostly along
$r$ and not $\theta$ from its URS value of 0.71 and UTS $\simeq$
0.04.  So (SPH,2,3) will strongly affect the shape of the
nanoparticle surface, as shown in Fig.~\ref{prf12}.

$u_L$ and $u_T$ take on simpler forms as $\eta$ becomes
larger.  For large $\eta$, the $u_L$ term has primarily
radial displacement, while the $u_T$ term corresponds
to displacement in the $\theta$ direction.

For the lowest modes, the situation is qualitatively different.
Consider (SPH,2,0) with $v_T/v_L = 0.5$.  Suppose to simplify
this discussion we normalize the displacement field so that
$(u|u) = 1$.   Then $(u_L|u_L) \simeq 0.14$.  However,
$(u_T|u_T) \simeq 1.85$.  So L3 for (SPH,2,0) is actually $\simeq$ -0.85,
making it ``ultra transverse'' by that measure.  It seems quite
odd that the $u_T$ term alone has a magnitude much greater than
that of the overall motion.  The resolution of this puzzle is
that $u_L$ and $u_T$ are not mutually orthogonal with respect to
the inner product of Eq.~\ref{norm}.  In fact, $(u_T|u_L)$
$\simeq$ -0.50.  According to the usual vector relation,
$\vec{a} \cdot \vec{b}$ = $\| \vec{a} \| \| \vec{b} \| cos \theta_{ab}$,
the ``angle'' between $u_L$ and $u_T$ is $\simeq$ 165 degrees
for the (SPH,2,0) mode.  This angle is nearly unchanged as
$v_T/v_L$ varies.  Thus, $u_L$ and $u_T$ are nearly antiparallel
vectors in the function space of vector fields within the
nanoparticle interior.  It can be said, then, that the functional
forms of $u_L$ and $u_T$ are actually relatively similar.  This
is a bit of a surprise since one is curl-free while the other is
divergence free.  This angle between $u_L$ and $u_T$ rapidly
approaches 90 degrees as $\eta$ increases (\textit{i.e.} for modes with
higher $n$).

As Fig.~\ref{SPH2} shows, the starkness of the contrast between     
SPH$_L$ and SPH$_T$ modes is at its best when $v_T/v_L$ is lower. For   
materials with high $v_T/v_L$ such as Si and Ge, FSM modes tend more to 
be mixtures of SPH$_L$ and SPH$_T$, especially at low $\eta$. But  
the concept of longitudinality is quite applicable to materials such as 
Au and Ag.                                                              

\section{High frequency mode classification}
\label{Highfreq}

The reason for the dichotomy of SPH modes as SPH$_T$ and SPH$_L$ can be
simply explained in the high frequency limit. Consider $\Delta_{\ell}$,
the 2 $\times$ 2 determinant in Eq.~\ref{twobytwo}, and its four matrix
elements. Note that $\xi / \eta$ = $v_T/v_L$. So at high frequency,
both $\eta$ and $\xi$ are large. In that case, $T_{11}$ and $T_{43}$
will be much larger than $T_{13}$ and $T_{41}$ because of their terms
including factors of $\eta^2$. Consequently, $\Delta_{\ell}$ is very
well approximated by $T_{11} T_{43}$. Since normal modes correspond to
zeroes of $\Delta_{\ell}$, it is clear that there will be two sets of
modes: those which are approximately zeroes of $T_{11}$ and $T_{43}$
respectively. The first group are SPH$_L$ and the second group are
SPH$_T$.

The roots of $T_{11}$ and $T_{43}$ are plotted
versus $\ell$ in Fig.~\ref{etavsl} with lines with crosses for
SPH$_L$ modes and lines with empty squares for SPH$_T$ modes.
The lines with full circles are the exact FSM mode frequencies.

There are three kinds of situations where we don't
expect this approximation to be valid: (1) for low $\eta$
(2) when it is not true that $\eta \gg \ell$ 
and (3) where longitudinal
($T_{11}$) and transverse ($T_{43}$) modes for a given $\ell$
are close -- \textit{i.e.} when the associated curves cross.  Except
in the previously mentioned places, the agreement between FSM
and our approximation is quite good.  The low $\eta$
situation corresponds specifically to the similar prefactors
of $j_{\ell}$ for T$_{11}$ and T$_{43}$ not being large.  It
is apparent that T$_{11}$ and T$_{43}$ can only be useful
as estimators of SPH$_L$ and SPH$_T$ mode frequencies when
$\eta \gg \ell$.  This is confirmed from inspection of the
lower right portion of Fig.~\ref{etavsl}.

For large $x$, $j_{\ell}(x) \simeq \sin(x-\ell\frac \pi 2) / x$.
Therefore, for large $\xi$ and $\eta$, the roots of $T_{11}$ can
be approximated by $\xi \simeq \ell \frac\pi 2 + (1+n_L) \pi$ and the
roots of $T_{43}$ by $\eta \simeq \ell \frac\pi 2 + n_T \pi$ where
$n_L \geq 0$ and $n_T \geq 0$ are integers.
These lead to remarkably compact approximate expressions for SPH$_L$    
and SPH$_T$ FSM frequencies in Hertz, respectively:                     
\begin{equation}
f \simeq \frac{v_L}{d} ( \frac{\ell}{2} + n_L + 1 )
\end{equation}
\begin{equation}
f \simeq \frac{v_T}{d} ( \frac{\ell}{2} + n_T )
\end{equation}
where $d$ = $2 R$. These expressions are very suggestive
of the formula for acoustic standing waves in a one dimensional
system of length $d$.
Table~\ref{def} shows the value of either
$n_L$ or $n_T$ for each mode.

The behaviour observed in Fig.~\ref{SPH2} becomes simple to
explain.  To a good approximation, SPH FSM modes are either
SPH$_L$ or SPH$_T$.  This approximation is considered here to be
good because it gives the right number of vibrational modes and
it predicts their frequency with a reasonable accuracy.

``Anti-crossing'' is observed in Fig.~\ref{SPH2} each time the
variation of the frequency of a SPH$_L$ mode crosses the one of
a SPH$_T$ mode. In Fig.~\ref{SPH2} there are two kinds of curves:
horizontal lines for SPH$_T$ modes and descending curves for
SPH$_L$ ones.  Then, each time these curves come together, an
anti-crossing pattern appears for the FSM solutions.  In
Fig.~\ref{etavsl}, FSM frequencies $\eta$ are plotted versus
$\ell$ for a sphere made of a material which has $v_T/v_L$ = 0.5.
Because the SPH$_L$ and SPH$_T$ approximation curves are plotted,
the anti-crossing patterns are clearly revealed.  The
continuation of Bessel functions to non-integer $\ell$ permits
relationships among modes for different integer $\ell$ to be
clearly seen.  This is preferable to the common practice of
joining modes on such a graph with hand-drawn straight lines.

\section{Discussion:}
\label{Discussion}

Normal elastic waves in a solid have a longitudinal acoustic
(LA) branch and two transverse acoustic (TA) branches.  However,
for FSM it seemed there are just two kinds: SPH and TOR.  By
classifying SPH modes into two kinds (\textit{i.e.} SPH$_L$ and
SPH$_T$), there are now three categories of modes, as we would
expect.

\begin{figure}[!ht]
	\includegraphics[width=\columnwidth]{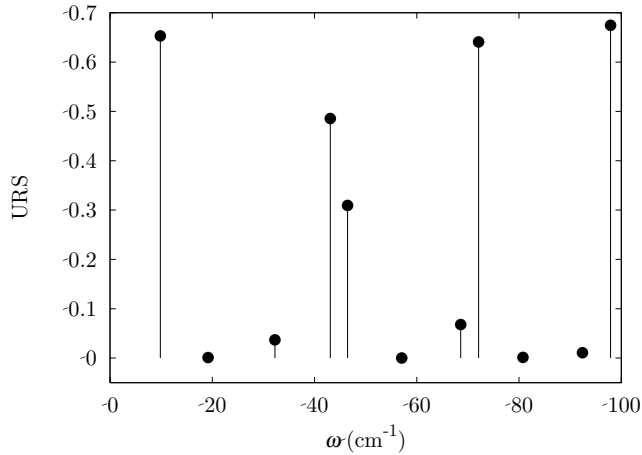}
  \caption{\label{AgSPH2}Mean squared radial surface displacement
  (URS) as a function of wavenumber for (SPH,$\ell=2$) modes of
  a 5 nm diameter silver nanoparticle ($v_T/v_L$ = 0.464).}
\end{figure}

We plot in Fig.~\ref{AgSPH2} the mean squared radial surface            
displacement (URS) at the surface of a 5~nm diameter silver             
nanoparticle for all SPH $\ell=2$ modes.                                
The magnitude of URS is in good agreement with the calculated   
Raman intensities\cite{bachelierPRB04}. (These calculations took into   
account the non-linear dispersion of acoustic phonons in silver. As a   
result, the calculated vibration wavenumbers do not match.) As discussed
before, (the SPH,2,0) mode is quite special even if we class it as a
SPH$_T$ mode. It changes the surface shape and therefore contributes
significantly to inelastic light scattering. Other harmonics contribute
significantly only when their URS is large and this in turn is very well
correlated to their SPH$_L$ nature as can be seen in Fig.~\ref{SPH2}.

Many experiments have observed peaks in Raman spectra attributed
to acoustic phonon vibrations of
silver
\cite{portales01a,portales01b,duval01,fujii91,portalesthesis,courty02,NeletASS04}
silicon
\cite{fujii96,saviotPRB03}
and CdS$_x$Se$_{1-x}$.
\cite{verma99,saviot96,ivanda03,irmer00,saviot98}
These studies have regularly succeeded in observing the (SPH,2,0) mode  
and the (SPH,0,0) mode. A number of studies have seen (SPH,0,$n$)       
with $n$ up to 4\cite{NeletASS04}.
However, there has never been a clear indication of   
Raman scattering from (SPH,2,1) even though there have been determined  
efforts to see it.

At the same time, (SPH$_L$,2,$n_L=0$) seems like a strong
candidate to have noticeable Raman scattering, since it has
strong radial surface motion as well as a strong $u_L$ component
that will give it stronger divergence in its interior.

It should be noted that $\ell = 0$ modes are always SPH$_L$.
That is why no full circles are plotted in Fig.~\ref{etavsl}
on the $T_{43}$ root curves at $\ell = 0$.  This has been the
source of many erroneous calculations in the past\cite{saviotPRE04}.

It is often
claimed\cite{tanaka93,tamura82,tamura83,ovsyuk96}
that modes with $n$ = 0 are ``surface
modes'' while modes with $n > 0$ are ``inner modes''.
The values of U2S in Tab.~\ref{def} show that
this is a misconception.  While this is true for $\ell$
= 0 and 1, for $\ell$ = 2, 3 and 4 it can be seen that
(SPH,$\ell$,1) has the strongest surface motion relative
to all (SPH,$\ell$,$n$).

Although Tab.~\ref{def} shows URS to be zero for (SPH,2,1),
the more precise value of $v_T/v_L$ where URS is zero is
0.488. URS for (SPH,2,1) is only near zero for $v_T/v_L$
close to 0.488.  However, URS remains small for (SPH,2,1)
for materials whose Poisson ratio is close to $\frac13$
which is true of many common materials. This contradicts
a widespread
misconception\cite{ohno00,lomnitz05,heyliger92}
that SPH FSM modes always have a radial displacement
component at the surface.

\begin{acknowledgments}
D. B. M. acknowledges support from the Natural Sciences and Engineering 
Research Council of Canada, the Okanagan University College             
Grant-in-Aid Fund and National Sun Yat-Sen University and thanks L. M.  
L. Murray and A. S. Laarakker for valuable scientific suggestions.      
\end{acknowledgments}

\bibliography{pr}
\end{document}